\documentclass[aps,prd,eqsecnum,preprint]{revtex4}
\usepackage[dvipdfmx]{color,graphicx}
\usepackage{amsfonts}
\usepackage{amsmath}
\usepackage{amssymb}
\usepackage{bm}

\newcommand{\bH}{\ensuremath{\bar{H}}}
\newcommand{\brho}{\ensuremath{\bar{\rho}}}

\begin{document}

{\baselineskip0pt
\rightline{\large\baselineskip16pt\rm\vbox to20pt{\hbox{OCU-PHYS-450}
            \hbox{AP-GR-131}
\vss}}%
}

%\thispagestyle{empty}

%<<<<<<<<<<<<< TITLE >>>>>>>>>>>>>>>%
\title{Can we remove the systematic error due to isotropic inhomogeneities?}

%<<<<<<<<<<<<< AUTHOR >>>>>>>>>>>>>>>%
\author{
Hiroyuki Negishi\footnote{Electronic address:negishi@sci.osaka-cu.ac.jp} and 
Ken-ichi Nakao\footnote{Electronic address:knakao@sci.osaka-cu.ac.jp}
}

\affiliation{
Department of Mathematics and Physics,
Graduate School of Science, Osaka City University, 
3-3-138 Sugimoto, Sumiyoshi, 
Osaka 558-8585, Japan
}
%<<<<<<<<<<<<< DATE >>>>>>>>>>>>>>>%

\begin{abstract}
Usually, we assume that there is no inhomogeneity isotropic in terms of our location in our universe.
This assumption has not been observationally confirmed yet in sufficient accuracy, and we need to consider the possibility that there are non-negligible large-scale isotropic inhomogeneities in our universe. 
The existence of large-scale isotropic inhomogeneities affects the determination of the cosmological parameters.
In particular, from only the distance-redshift relation,  we can not distinguish the inhomogeneous isotropic universe  model from the homogeneous isotropic one, because of the ambiguity in the cosmological parameters.
In this paper, in order to avoid such ambiguity, we consider three observables,
the distance-redshift relation, the fluctuation spectrum of the cosmic microwave background radiation(CMBR) and the scale of the baryon acoustic oscillation(BAO), and compare these observables in two universe models; One is  
the inhomogeneous isotropic universe model with the cosmological constant and the other is  
the homogeneous isotropic universe model with the dark energy other than the cosmological constant.
We show that these two universe models can not predict the same observational data of all three 
observables but the same ones of only two of three, as long as the perturbations are adiabatic.  
In principle, we can distinguish the inhomogeneous isotropic universe 
from the homogeneous isotropic one through appropriate three observables, if the perturbations are adiabatic.

\end{abstract}

\maketitle

%\tableofcontents 

\vskip1cm

%%%%%%%%%%%%%%%%%%%%%%%
%%%%%%%%%%%%%%%%%%%%%%%
\section{Introduction}\label{Sec1}
%%%%%%%%%%%%%%%%%%%%%%%
%%%%%%%%%%%%%%%%%%%%%%%

Usually, the modern physical cosmology adopts the Copernican principle which states that we do not live in the privileged domain in the universe. 
The Copernican principle and the observed high isotropy of the CMBR provide the high homogeneity and isotropy of our universe in the globally averaged sense. 
By contrast, if we remove the Copernican principle, the isotropy around us does not necessarily imply the homogeneity of our universe.
The Copernican principle can not be directly confirmed by observations since in order to do so we have to move to the other clusters of galaxies from our galaxy.
Hence there is the possibility that there are large-scale isotropic inhomogeneities in our universe. 
The existence of isotropic inhomogeneities around us affects the determination of the cosmological parameters.

The universe model with large-scale isotropic inhomogeneities has been studied in the context of the scenario to explain the observed distance-redshift relation without introducing dark energy components within the framework of general relativity. 
There are several severe observational constraints on the scenario without dark energy\cite{Tomita:2000jj,Tomita:2001gh,Celerier:1999hp,Iguchi:2001sq,Yoo:2008su,Bull:2012zx,Clifton:2008hv,Vanderveld:2006rb,Yoo:2010qn},
in particular, these universe models are constrained by observations of the kinetic Sunyaev-Zeldovich effect\cite{Zhang,Zibin:2010a,Zibin:2011ma}; 
The scenario with adiabatic isotropic inhomogeneities has already been ruled out,  
even though both growing and decaying modes are assumed to exist. 
On the other hand, the scenario with non-adiabatic isotropic inhomogeneities has not been ruled out yet,  
although there is an argument on whether the initial condition is contrived.

Even if there are dark energy components,
not so large isotropic inhomogeneities may exist and significantly affect observational results\cite{Romano:2010nc,Romano:2011mx,Marra:2010pg,Sinclair:2010sb,Valkenburg:2013qwa,Valkenburg:2012td,deLavallaz:2011tj,Valkenburg:2011ty,Marra:2012pj,Tokutake:2016hod,Negishi:2015oga}.
Since our observation is confined on a past light cone, a spatially more distant event we observe 
occurred a longer time ago.
If the universe is homogeneous and isotropic, the temporal evolution of the universe is revealed by observing events of various distances.
By contrast, if the universe is inhomogeneous and isotropic, observational data contain the information 
about not only the temporal evolution of the universe but also its spatial inhomogeneities.
An intrinsic degeneracy between temporal evolution and isotropic inhomogeneities around 
us may cause systematic errors. 
Denoting the total energy density and the total pressure of dark energy components 
by $\rho_{\rm d}$ and $p_{\rm d}$, respectively, its equation of state is given by 
\begin{equation}
p_{\rm d}=w\rho_{\rm d} \label{EOS}
\end{equation}
with $\rho_{\rm d}>0$, $w$ is a function less than $-1/3$. The case of $w=-1$ 
corresponds to the cosmological constant. 
We can not distinguish the inhomogeneous isotropic universe model with 
the cosmological constant from the homogeneous isotropic universe model 
with dark energy of $w\neq -1$, if we have the observational date of the distance-redshift 
relation only; isotropic inhomogeneities may cause systematic errors on the 
amount of dark energy and its equation of state. Some authors studied systematic errors due to isotropic inhomogeneities and have evaluated the magnitude of them\cite{Marra:2010pg,Valkenburg:2012td,Romano:2010nc,Romano:2011mx,Sinclair:2010sb,Valkenburg:2013qwa,deLavallaz:2011tj,Valkenburg:2011ty,Marra:2012pj,Tokutake:2016hod,Negishi:2015oga}. 
However, nobody has not shown how to remove the systematic errors caused by isotropic inhomogeneities. 
It is just the main purpose of the present paper. 

The homogenous isotropic universe model is often called the Friedmann-Lema\^{\i}tre-Robertson-Walker (FLRW) universe model, and hereafter we call so.

In this paper, we study whether we can distinguish the inhomogeneous isotropic universe model with the cosmological constant from the FLRW universe model with dark energy of $w\neq -1$ and remove the systematic error due to isotropic inhomogeneities, if we use multiple observables: 
 the distance-redshift relation, the fluctuation spectrum of the CMBR and the scale of the BAO in the distribution of galaxies.
In other words, we investigate whether there is an inhomogeneous isotropic universe model  whose distance-redshift relation, fluctuation spectrum of the CMBR and BAO scale are identical with those of the FLRW universe model.
We assume that the inhomogeneous isotropic universe model is filled with non-relativistic matter which is cold dark matter(CDM) and baryonic matter and a positive cosmological constant. 
Furthermore, we restrict ourselves to the case that the amplitude of isotropic inhomogeneities is so small that they can be treated by the linear perturbation approximation of the flat FLRW universe model, and the scale of isotropic inhomogeneities are larger than the BAO scale but should be smaller than the present horizon scale.
If we can find such an inhomogeneous isotropic universe model, it is impossible to distinguish these two universe models from each other, otherwise we can.

The organization of this paper is as follows.
In Sec.~II, we derive the basic equations for the inhomogeneous isotropic universe model.
In Sec.~III, we derive the null geodesic equations which are used to construct the inhomogeneous isotropic universe model from observables given from the FLRW universe model.
In Sec.~IV, we show expressions of observables in the inhomogeneous isotropic universe model.
In Sec.~V, we give the observables in the FLRW model and derive conditions to determine the inhomogeneous isotropic universe model.
We explain the numerical procedure and the numerical result in Sec.~VI. 
Finally, Sec.~VII is devoted to the summary and discussion.

In this paper, we adopt the sign conventions of the metric and Riemann tensor of Ref.\cite{wald} and 
the geometrized unit in which the speed of light and Newton's gravitational constant are one.

%%%%%%%%%%%%%%%%%%%%%%%
%%%%%%%%%%%%%%%%%%%%%%%
\section{Inhomogeneous isotropic universe model}\label{Sec2}
%%%%%%%%%%%%%%%%%%%%%%%
%%%%%%%%%%%%%%%%%%%%%%%
As mentioned in Sec.\ref{Sec1}, the inhomogeneous isotropic universe model is described by the flat FLRW universe model with isotropic linear perturbations. 
By adopting the synchronous comoving gauge, the infinitesimal world interval is written in the form, 
\begin{eqnarray}
ds^2&=&-dt^2+a^2(t)\left[ (1+X(t, r)) \delta_{ij} +\partial_i \partial_j Y(t,r)  \right] dx^i dx ^j \cr \cr
&=&-dt^2+a^2\left[ (1+X+\partial _r^2 Y) dr^2 +\left( 1+X+\frac{1}{r}\partial _r Y \right) r^2 d\Omega ^2  \right],
\label{metric} 
\end{eqnarray}
where $a(t)$ is the scale factor scaled so as to be unity at the present time $t=t_0$ and $d \Omega^2$ is the line element of the unit 2-sphere.

We assume that this universe model is filled with non-relativistic matter and the cosmological constant $\Lambda$. 
The stress-energy tensor of the non-relativistic matter  is given by
\begin{eqnarray}
T_{\mu\nu}=\bar{\rho}_{\rm m}(t)(1+\Delta (t,r))u_\mu u_\nu ,
\label{T_ab} 
\end{eqnarray}
where $\bar{\rho}_{\rm m}$, $\Delta$ and $u_\mu$ are the energy density of the background,  the density contrast and the 4-velocity, respectively.
The coordinate system is chosen so that the components of the  4-velocity is given by $u_\mu=(-1,0,0,0)$.

The Einstein equations lead to the Friedmann equation for the background;  
\begin{eqnarray}
H^{2}(a):=\left(\frac{1}{a}\frac{da}{dt}\right)^2=\frac{8\pi \brho _{\rm m0}}{3a^{3}}+\frac{\Lambda}{3} , 
\label{Eeq_b} 
\end{eqnarray}
where $\bar{\rho}_{\rm m0}$ is the background energy density at $t=t_0$. 
Denoting the present value of $H$ by $\bH_0$, Eq.~(\ref{Eeq_b}) is rewritten in the form
\begin{equation}
H^2=\bH_0^2\left(\frac{\Omega_{\rm m}}{a^3}+\Omega_\Lambda\right),
\label{H-eq}
\end{equation}
where
\begin{equation}
\Omega_{\rm m}=\frac{8\pi\bar{\rho}_{\rm m0}}{3\bH_0^2},~~~~{\rm and}~~~~
\Omega_\Lambda=\frac{\Lambda}{3\bH_0^2}.
\end{equation}
The Einstein equations lead to the equations for the linear perturbations;   
\begin{eqnarray}
\dot{X}=0, 
\label{Eeq_p0}
\end{eqnarray}
\begin{eqnarray}
X-a^2 \ddot{Y}-3a\dot{a}\dot{Y}=0,
\label{Eeq_p1}
\end{eqnarray}
\begin{eqnarray}
\partial _i \partial ^i \dot{Y} =-2 \dot{\Delta},
\label{Eeq_p2}
\end{eqnarray}
\begin{eqnarray}
\ddot{\Delta} +2H\dot{\Delta} -4\pi \frac{\brho _{\rm m0}}{a^3}\Delta=0,  
\label{Eeq_p3}
\end{eqnarray}
where a dot denotes a partial differentiation with respect to $t$.

The general solution of Eq.~(\ref{Eeq_p3}) is represented 
by the linear superposition of the growing factor $D_{+}(t)$ and the decaying factor $D_{-}(t)$, which are defined as 
\begin{equation}
D_{+}(t):=\bH _0^2\left( H(a)\int ^{a}\frac{1}{b^{3}H^{3}(b)} db \right) ~~~~~{\rm and}~~~~~D_{-}(t):=\frac{H(a)}{\bH _0}.
\label{Dpm-sol}
\end{equation}
Hereafter, we assume that the decaying mode does not exisit, since this assumption is consistent with 
the inflationary universe scenario.  
Thus, we have 
\begin{eqnarray}
\Delta (t,r)=f(r) D_{+}(t) ,
\label{delta-f}
\end{eqnarray}
where $f(r)$ is an arbitrary function of the radial coordinate $r$. 

From Eqs.~(\ref{Eeq_p0}) -- (\ref{Eeq_p2}), $X$ and $Y$ are expressed by 
\begin{eqnarray}
X=-h(r),
\label{A-h}
\end{eqnarray}
\begin{eqnarray}
Y=-\frac{h(r)}{\bH _0^2}D_+ +Y_0(r),
\label{A-h}
\end{eqnarray}
where $h(r)$ and $Y_0(r)$ are arbitrary functions of the radial coordinate $r$.
A gauge freedom remains in Eq.~(\ref{metric}), so we fix the gauge so that $Y_0(r)=0$.
Through Eq.~(\ref{Eeq_p2}), we have
\begin{eqnarray}
\Delta = \frac{1}{2\bH _0^2} D_+\partial _i \partial ^i h.
\label{delta-h}
\end{eqnarray}

%%%%%%%%%%%%%%%%%
%%%%%%%%%%%%%%%%%
\section{Null geodesics in inhomogeneous isotropic universe model}
%%%%%%%%%%%%%%%%%
%%%%%%%%%%%%%%%%%
In order to calculate some observable, we consider a past-directed radial null geodesic which emanates from the observer.
We assume that the observer in the inhomogeneous isotropic universe model stays at the symmetry center $r=0$, so that the observer recognizes the universe to be isotropic. 
By virtue of the isotropy in terms of the observer, both $k^{\theta}$ and $k^{\phi}$ should vanish.
One of the non-trivial components of the geodesic equations is given by
\begin{equation}
\frac{d}{d\lambda}k^t +\left[ H -\frac{\dot{D_+}}{2\bH _0^2}\partial _r^2h\right] (k^t)^2=0,
\label{g-eq-1}
\end{equation}
where $\lambda$ is the affine parameter. Equation (\ref{g-eq-1}) determines $k^t$, whereas 
the null condition determines $k^r$ in the manner
\begin{equation}
k^{r}=-\frac{1}{a}\left( 1+ \frac{h}{2}+\frac{D_+}{2\bH _0^2}\partial _r ^2 h\right) k^{t}.
\label{null-cond-1} 
\end{equation}
Then we have equations for $t$ and $r$ as 
\begin{eqnarray}
\frac{dt}{d\lambda}&=&k^t,  \label{g-eq-for-t} \\
\frac{dr}{d\lambda}&=&k^r. \label{g-eq-for-r}
\end{eqnarray}

The redshift $z$ for the observer is given by  
\begin{equation}
1+z=\frac{(k^{\mu}u_{\mu})|_{{\rm s}}}{(k^{\mu}u_{\mu})|_{{\rm o}}} =k^t,
\label{redshift-1}
\end{equation}
where subscripts s and o mean the quantities evaluated at the source and the observer, respectively, 
and we have chosen the affine parameter so that $-(k^\mu u_\mu)_{\rm o}$ is unity. 

We rewrite the equations for the radial null geodesic in the forms appropriate for later analyses.  
Equation~(\ref{g-eq-1}) is rewritten in the form, 
\begin{equation}
\frac{1}{k^t}\frac{d k^t}{dz}=-\left(H-\frac{1}{2\bH _0^2}\dot{D_+}\partial _r^2 h\right)\frac{dt}{dz}.
\label{g-eq-2}
\end{equation}
From 
Eqs.~(\ref{redshift-1}) and (\ref{g-eq-2}), 
we have
\begin{equation}
\frac{1}{1+z}+\left(H-\frac{1}{2\bH _0^2}\dot{D_+}\partial _r^2 h\right)\frac{dt}{dz}=0.
\label{basic-eq-1}
\end{equation}
From the null condition (\ref{null-cond-1}), we have
\begin{equation}
\frac{dr}{dz}= -\frac{1}{a}\left( 1+ \frac{h}{2}+\frac{D_+}{2\bH _0^2}\partial _r ^2 h\right) \frac{dt}{dz}.
\label{basic-eq-2} 
\end{equation} 
We express the radial null geodesic as a function of $z$; 
\begin{eqnarray}
t&=&\bar{t}(z)+\delta t(z),  \label{t(z)} 
\label{t}
\end{eqnarray}
\begin{eqnarray}
r&=&\bar{r}(z)+\delta r(z), \label{r(z)}
\label{r}
\end{eqnarray}
where the quantities with a horizontal bar represent the background solution. 
From Eqs.~(\ref{basic-eq-1}) and (\ref{basic-eq-2}), we see that 
the background solutions $\bar{t}(z)$ and $\bar{r}(z)$ satisfy  
\begin{eqnarray}
\frac{d\bar{t}}{dz}=-\frac{1}{(1+z)\bar{H}}, 
\label{g-eq_bt}
\end{eqnarray}
\begin{equation}
\frac{d\bar{r}}{dz}=\frac{1}{\bar{H}},
\label{g-eq_br}
\end{equation}
where
$$
\bar{H}=\bH_{0}\sqrt{\Omega _{\rm m}(1+z)^{3}+\Omega _{\Lambda}}. 
$$
Equation~(\ref{basic-eq-1}) leads to  the equation for the linear perturbations as
\begin{equation}
\frac{d\delta t}{dz}=-\frac{1}{\bar{H}}\frac{d\bar{H}}{dz}\delta t+\frac{1}{2\bH _0^2}\frac{dD_+}{dz}\frac{d}{dz}\left( \bar{H}\frac{dh}{dz}\right) ,
\label{basic-1}
\end{equation}
whereas Eq.~(\ref{basic-eq-2}) leads to 
\begin{equation}
\frac{d\delta r}{dz}= \frac{h}{2\bH}+\frac{D_+}{2\bH _0^2}\frac{d}{dz}\left( \bar{H}\frac{dh}{dz}\right) -\delta t-(1+z)\frac{d\delta t}{dz},
\label{basic-2}
\end{equation}
where we have used Eqs.~(\ref{g-eq_bt}) and (\ref{g-eq_br}). 

The boundary conditions for Eqs.~(\ref{g-eq_bt})--(\ref{basic-2}) 
are given by 
\begin{equation}
\bar{t}(0)=t_0~~~~{\rm and}~~~~\bar{r}(0)=\delta t (0)=\delta r (0)=0. 
\label{b-condi} 
\end{equation}
If the inhomogeneous isotropic universe model is completely fixed, 
Eqs.~(\ref{basic-1}) and (\ref{basic-2}) determine $\delta t$ and $\delta r$.

%%%%%%%%%%%%%%%%%
%%%%%%%%%%%%%%%%%
\section{Observables}
%%%%%%%%%%%%%%%%%
%%%%%%%%%%%%%%%%%
In this Section, we show observables of our interest in the inhomogeneous isotropic universe model.

%%%%%%%%%%%%%%%%%
\subsection{Angular diameter distance-redshift relation}
%%%%%%%%%%%%%%%%%
By observing many supernovae, we have the correlation between the luminosity distance $d_{\rm L}$ and redshift $z$.
In this paper, we treat the angular diameter distance $d_{\rm A}$ instead of $d_{\rm L}$;
$d_{\rm L}$ is obtained from $d_{\rm A}$ through
\begin{eqnarray}
d_{\rm L}=(1+z)^2 d_{\rm A}.
\end{eqnarray}
In the inhomogeneous isotropic universe model, 
the angular diameter distance from some light source to the observer 
is equal to the areal radius at which the light is emitted; 
\begin{eqnarray}
d_{\rm A}(z)&=&a \Bigl[1+\frac{1}{2}X +\frac{1}{2r}\partial _r Y \Bigr] r \Bigl|_{t=t(z),~r=r(z)}, \cr
&&\cr
&\approx&a(\bar{t})\bar{r}+a(\bar{t})\left[ \delta r -\left( \frac{1}{2}h(\bar{r}) +\frac{\bar{H}D_+(\bar{t})}{2\bar{r}\bH _0^2}\frac{d h(\bar{r})}{dz}-\bH \delta t \right) \bar{r}\right] .
\label{condi-1}
\end{eqnarray}
Equation (\ref{condi-1}) 
implies that the angular diameter distance-redshift relation depends on two parameters 
$\brho_{\rm m0}$, $\Lambda$ and one arbitrary function $h$, since $\bar{t}, \bar{r},\delta t$ and $\delta r$ are determined by the geodesic equations (\ref{g-eq_bt})--(\ref{basic-2}), if we give $\brho_{\rm m0}$, $\Lambda$ and $h$

%%%%%%%%%%%%%%%%%
\subsection{Cosmic microwave background radiation}\label{CMBR}
%%%%%%%%%%%%%%%%%
Anisotropies of the CMBR are important observables.
They come from the anisotropies of the last scattering surface and the Integrated Sachs-Wolfe(ISW) effect.
In this paper, we are interested in the high multipoles of the anisotropies of the CMBR for which the ISW effect is not important, since there is a slight information in low multipoles because of their large  cosmic variance.
Hereafter, we ignore the ISW effect, and follow \cite{Clarkson-Regis} to calculate the anisotropies of the CMBR in the inhomogeneous isotropic universe model.

Hereafter, we adopt the following assumptions;
\begin{itemize}
\item  The inhomogeneous isotropic universe model well agrees with the background universe model in the vicinity of the last scattering surface(LSS);
\item The background universe model well agrees with the Einstein-de Sitter(EdS) universe model at the LSS;
\item  The non-relativistic matter is composed of baryonic matter and CDM whose background energy densities are represented by $\bar{\rho} _{\rm b}$ and $\bar{\rho}_{\rm CDM}=\bar{\rho}_{\rm m}-\bar{\rho}_{\rm b}$ respectively;
\item  The primordial density fluctuations is adiabatic and its power spectrum $P(k)$ is characterized by an amplitude $A_0$ and a spectral index $n_s$;
\begin{eqnarray}
P(k)=A_0\left( \frac{k}{k_0}\right) ^{n_s},
\label{Pk}
\end{eqnarray}
where $k_0=0.05 {\rm Mpc} ^{-1}$.
\end{itemize}
If these assumptions are valid, the CMBR angular power spectrum $C_l$ is given by
\begin{eqnarray}
C_l =S^{-2} C_{S^{-1}l}^{\rm (EdS)} ~~~~{\rm for}~  l \gg 1,
\label{cl}
\end{eqnarray}
where $C_l^{\rm (EdS)}$ is the CMBR angular power spectrum of the EdS universe model filled with baryonic matter and CDM whose energy densities at the LSS are the same as $\brho _{\rm b}^{\rm LSS}$ and $\brho _{\rm CDM}^{\rm LSS}$, respectively,
where superscripts LSS mean quantities at the LSS,
 and the primordial power spectrum is the same as that given by Eq.(\ref{Pk}), and
\begin{eqnarray}
S=\frac{d_{\rm A}(z_{\rm LSS})}{d_{\rm A}^{\rm (EdS)}(z_{\rm LSS})},
\label{S}
\end{eqnarray}
where $z_{\rm LSS}$ is the redshift at the LSS, and $d_{\rm A}^{\rm (EdS)}(z)$ is the angular diameter distance at $z$ in the EdS universe model.
We choose the present temperature of the CMBR to be $2.725 {\rm K}$ and the spectral index $n_s$ to be $0.9667$.
Equation (\ref{cl}) means that the CMBR angular power spectrum of the inhomogeneous isotropic universe model depends on $\brho _{\rm m}^{\rm LSS}$, $\brho _{\rm b}^{\rm LSS}$ and $A_0$ which characterize $C_{l}^{\rm (EdS)}$ and $d_{\rm A}(z_{\rm LSS})$.
Usually, $C_l$ depends on the function $h$ in the inhomogeneous isotropic universe model, but Eq.~(\ref{cl}) depends on a finite parameters, since we ignore the ISW effect.

%%%%%%%%%%%%%%%%%
\subsection{Baryon Acoustic Oscillation}
%%%%%%%%%%%%%%%%%
Large scale redshift surveys of galaxies tell us the BAO scale\cite{Eisenstein:2005su,Percival et al.(2010),Blake et al.(2011)}.
Observables related to the BAO scale are $\Delta \theta _{\rm BAO}$ which is the angular diameter of the BAO scale and $\Delta z _{\rm BAO}$ which is the BAO scale in the redshift space.
Most papers on the BAO observations quote  
\begin{eqnarray}
d_{\rm z}(z)=\left( \Delta \theta _{\rm BAO}^2 \frac{\Delta z _{\rm BAO}}{z}\right) ^{1/3}.
\label{dz}
\end{eqnarray}
In order to calculate $d_{\rm z}$, first of all, we need the BAO scale at the decoupling time, 
which is denoted by $L_{\rm BAO}^{\rm (dec)}$.
In the inhomogeneous isotropic universe model, 
 the isotropic density perturbation has been assumed to be composed of the only growing mode which is so small that 
 the liner perturbation approximation is valid until  the present time.
We assume that the ratio of baryonic matter and CDM is everywhere constant, so that we regard that 
the inhomogeneous and isotropic universe model is almost homogeneous and isotropic at the decoupling time 
and $L_{\rm BAO}^{\rm (dec)}$ is everywhere constant.
This assumption is the same as that in ref.~\cite{Zumalacarregui-Bellido-Lapuente}.
The values of the BAO scale in transverse and radial directions, $L_{\rm BAO}^{\rm T}$ and $L_{\rm BAO}^{\rm R}$, at an event on the past light cone are respectively given by
\begin{eqnarray}
L_{\rm BAO}^{\rm T}(z) &:=& \int _{\theta}^{\theta +\Delta \theta _{\rm BAO}}\sqrt{g_{\theta \theta}(t(z),r(z))}d\theta, \cr \cr
&\approx& \frac{\sqrt{g_{\theta \theta}(t(z),r(z))}}{\sqrt{g_{\theta \theta}(t _{\rm dec},r(z))}} L_{\rm BAO}^{\rm (dec)}, \cr \cr
&=&a(\bar{t})\left( 1+\bH (z)\delta t(z) -\frac{1}{2}h(z)-\frac{D_+(\bar{t}(z))}{2\bar{r}(z)\bH _0^2}\partial _r h(\bar{r}(z))\right) \frac{L_{\rm BAO}^{\rm (dec)}}{a(t_{\rm dec})},
\label{L_BAO-T}
\end{eqnarray}
\begin{eqnarray}
L_{\rm BAO}^{\rm R}(z) &:=&  \int _{r(z)}^{r(z+\Delta z _{\rm BAO}) }\sqrt{g_{rr}(t(z),x)}dx, \cr \cr
&\approx& \frac{\sqrt{g_{rr}(t(z),r(z))}}{\sqrt{g_{rr}(t _{\rm dec},r(z))}} L_{\rm BAO}^{\rm (dec)}, \cr \cr
&=&a(\bar{t})\left( 1+\bH (z)\delta t (z)-\frac{1}{2}h(z)-\frac{D_+(\bar{t}(z))}{2\bH _0^2 }\partial _r^2 h(\bar{r}(z))\right) \frac{L_{\rm BAO}^{\rm (dec)}}{a(t_{\rm dec})}.
\label{L_BAO-R}
\end{eqnarray}
where $t_{\rm dec}$ is the decoupling time, $a(t_{\rm dec})$ is scale factor of decoupling time, and in the second equality in the above equations we have used the fact that the BAO scale is comoving in 
the gauge adopted here  and the scale of isotropic inhomogeneities are larger than the BAO scale.
From Eq.~(\ref{L_BAO-T}), we have
\begin{eqnarray}
\Delta \theta _{\rm BAO} (z)&=&\frac{L_{\rm BAO}^{\rm T}(z)}{d_{\rm A}(z)}, \cr \cr
&=& \frac{L_{\rm BAO}^{\rm (dec)}}{d_{\rm A}}\frac{1}{a(t_{\rm dec})(1+z)}\left( 1+\bH \delta t-\frac{1}{2}h-\frac{\bH D_+}{2\bar{r}\bH _0^2}\frac{dh}{dz}\right).
\label{delta-theta}
\end{eqnarray}
$\Delta z _{\rm BAO}$ is obtained from integrating geodesic equations (\ref{basic-eq-1}) -- (\ref{r}) 
\begin{eqnarray}
\Delta z _{\rm BAO} (z) &=& \int _{r(z)}^{r(z+\Delta z _{\rm BAO})}\frac{dz}{dr}dr, \cr \cr
&\approx&(1+z)\left( H-\frac{1}{2\bH_0^2}\dot{D}_+\partial _r^2h\right) L_{\rm BAO}^{\rm R}(z), \cr \cr
&=& \frac{L_{\rm BAO}^{\rm (dec)}}{a(t_{\rm dec})}\bH \left[ 1+\left( \bH-(1+z)\frac{d\bH}{dz} \right)\delta t -\frac{1}{2}h \right. \cr \cr
& &{}\left. -\frac{\bH}{2}\left( (1+z)\frac{dD_+}{dz}+D_+\right) \frac{d}{dz}\left( \bH\frac{dh}{dz}\right) \right].
\label{delta-z}
\end{eqnarray}
where in the second equality in the above equations we have used the fact that the BAO scale is comoving and the scale of isotropic inhomogeneities are larger than the BAO scale.
Thus, we obtain
\begin{eqnarray}
d_{\rm z}(z)&=& \frac{L_{\rm BAO}^{\rm (dec)}}{a(t_{\rm dec})}\left( \frac{1}{d_{\rm A}^2}\frac{\bH}{z(1+z)^2}\right) ^{1/3} \left[ 1+\bH \delta t -\frac{1}{3}(1+z)\frac{d\bH}{dz}\delta t-\frac{1}{2}h-\frac{\bH D_+}{3\bar{r}\bH_0^2}\frac{dh}{dz}\right. \cr \cr
& &{}\left. -\frac{\bH}{6\bH_0^2}\left( (1+z)\frac{dD_+}{dz}+D_+\right) \frac{d}{dz}\left( \bH\frac{dh}{dz}\right) \right] .
\label{dz-inhomo}
\end{eqnarray}
Equation (\ref{dz-inhomo}) means that $d_{\rm z}$ in the inhomogeneous isotropic universe model depends on $a(t_{\rm dec})$, $\brho _{\rm m0}$, $\brho _{\rm b0}$, $\Lambda$ and $h$.
$L_{\rm BAO}^{\rm (dec)}$ is determined by $\brho _{\rm b}(t_{\rm dec})$ and $\brho _{\rm m}(t_{\rm dec})$, where
\begin{eqnarray}
\brho _{\rm m}(t_{\rm dec}) = \frac{\brho_{\rm m0}}{a^3(t_{\rm dec})}
\label{rho_dec}
\end{eqnarray}
\begin{eqnarray}
\brho _{\rm b}(t_{\rm dec}) = \frac{\brho_{\rm b0}}{a^3(t_{\rm dec})}
\label{rhob_dec}
\end{eqnarray}

%%%%%%%%%%%%%%%%%
%%%%%%%%%%%%%%%%%
\section{How to construct the inhomogeneous isotropic universe model}
%%%%%%%%%%%%%%%%%
%%%%%%%%%%%%%%%%%
As mentioned in Sec \ref{Sec1}, we study whether we can distinguish the inhomogeneous isotropic universe model with the cosmological constant from the FLRW universe model with dark energy other than the cosmological constant, if we consider multiple observables.

We choose the FLRW universe model as follows;
The FLRW universe model is filled
with non-relativistic matter which is composed of baryonic matter and CDM and dark energy 
whose $w$ in the equation of state (\ref{EOS}) is written in the form
\begin{eqnarray}
w=\sum_{n=0}^\infty w_n(1-\hat{a})^n, \label{w-def} 
\end{eqnarray}
where $\hat{a}$ is the scale factor of the FLRW universe model, and $w_n$ is constant and $w_0 < -1/3$;  
We assume that $w_n=0$ for $n\geq2$;
Furthermore, for simplicity, we assume that the FLRW universe model has flat space, $k=0$. 
The equation of state of this form was studied in the cosmological context by M.~Chevallier and D.~Polarski\cite{Chevallier:2000qy}.
This FLRW universe model is characterized by five parameters, Hubble constant $\hat{H}_0$, $w_{0}$, $w_1$, the present value of energy density of baryonic matter $\hat{\rho}_{\rm b0}$ and of dark energy $\hat{\rho}_{\rm d0}$.

In accordance with the standard scenario of the structure formation,
we assume the adiabatic primordial density fluctuations in the FLRW universe model, whose power spectrum $\hat{P}(k)$ is given by
\begin{eqnarray}
\hat{P}(k)=\hat{A}_0\left( \frac{k}{k_0}\right) ^{\hat{n}_s},
\label{Pkf}
\end{eqnarray}
where $\hat{A}_0$ and $\hat{n}_s$ are an amplitude and a spectral index,  respectively.

We investigate whether we can construct the inhomogeneous isotropic universe model which satisfies the following conditions;
on the angular diameter distance-redshift relation
\begin{eqnarray}
d_{\rm A}(z)=\hat{d}_{\rm A}(z) ~~~~{\rm for}~  0<z<2;
\label{da-con}
\end{eqnarray}
on the angular power spectrum of the CMBR 
\begin{eqnarray}
C_l=\hat{C}_{l} ~~~~{\rm for}~  l \gg 1;
\label{cl-con}
\end{eqnarray}
on the averaged angular scale of the BAO 
\begin{eqnarray}
d_{\rm z}(z)=\hat{d}_{\rm z}(z) ~~~~{\rm at}~  z=0.2, ~ 0.35,
\label{dz-con}
\end{eqnarray}
where characters with a hat denote quantities of the FLRW universe model.
Note that the condition on the angular diameter distance-redshift relation is restricted in the domain $0<z<2$. 
This is because we do not have any observational data of the distance-redshift relation in the domain of $z\geq 2$.

For convenience, we define a new variable
\begin{eqnarray}
v(z)=\bar{H}(z)\frac{dh(\bar{r}(z))}{dz}.
\label{v}
\end{eqnarray}
The condition (\ref{da-con}) together with Eq.~(\ref{condi-1}) gives us the relation between $h$, $v$, $\delta t$ and $\delta r$,
\begin{eqnarray}
\delta r =(1+z)\delta d_{\rm A}+ \frac{D_+}{2\bH_0}v+\left( \frac{1}{2}h-\bar{H}\delta t\right) \bar{r},
\label{deltar}
\end{eqnarray}
where 
\begin{equation}
\delta d_{\rm A}(z) =\hat{d}_{\rm A}(z)-\frac{\bar{r}}{1+z}.
\end{equation}
By substituting Eq.~(\ref{deltar}) into Eq.~(\ref{basic-2}), we eliminate $\delta r$ from Eq.~(\ref{basic-2}) and obtain the differential equation for $\delta t$.
By eliminating $d\delta t/dz$ from Eq.~(\ref{basic-1}), we obtain the differential equation for $v$.
As a result, we obtain the following system of differential equations to determine the 
inhomogeneous isotropic universe model which satisfies the condition (\ref{da-con});
\begin{eqnarray}
\frac{d h}{dz}&=& \frac{v}{\bar{H}},
\label{eq-1} \\
&& \cr
\frac{d\delta t}{dz}&=&
\frac{{\cal N}\left( \delta t,v,z\right)}{{\cal D}(z)},
\label{eq-2} \\
&&\cr
\frac{d v}{dz}&=& {\cal V}\left( \delta t,v, z\right),
\label{eq-3} 
\end{eqnarray}
where
\begin{eqnarray}
{\cal D}(z)&=&\bar{r}\bar{H}-(1+z), \\
&&\cr
{\cal N}\left(\delta t,v,z\right) &=& \frac{d}{dz}\left[ (1+z)\delta d_{\rm A}\right] +\left(\frac{1}{2\bH_0^2}\frac{dD_+}{dz}+\frac{\bar{r}}{2\bar{H}} \right) v -\bar{r}\frac{d\bar{H}}{dz}\delta t ,\\
&& \cr
{\cal V}&=&
\frac{2\bH_0^2}{\frac{dD_+}{dz}}\left[ \frac{{\cal N}\left( \delta t,v,z\right)}{{\cal D}(z)}+\frac{1}{\bar{H}}\frac{d\bH}{dz} \delta t\right] .
\end{eqnarray}

The boundary conditions for Eqs.~(\ref{eq-1})--(\ref{eq-3}) are given as follows. 
The boundary condition on $\delta t$ is given by Eq.~(\ref{b-condi}). 
Imposing $C^1$ regularity for $h$ at $r=0$, i.e., $\partial_r h |_{z=0}=0$, 
we obtain
\begin{eqnarray}
v|_{z=0}=0.  
\label{bound_v} 
\end{eqnarray} 
The value of $h|_{z=0}$ can be made zero by a rescaling of the coordinate.
Hence we impose  
\begin{eqnarray}
h|_{z=0}=0. 
\label{bound_h} 
\end{eqnarray} 

In the next section, we explain how to solve  Eqs.~(\ref{eq-1})--(\ref{eq-3}) 
under the boundary conditions (\ref{b-condi}), (\ref{bound_v}) and (\ref{bound_h}). 
We numerically solve these equations and investigate whether the obtained 
inhomogeneous isotropic universe model satisfies the conditions (\ref{cl-con}) and (\ref{dz-con}).

%%%%%%%%%%%%%%%%%%%%%
%%%%%%%%%%%%%%%%%%%%%
\section{numerical procedure and result}\label{Sec6}
%%%%%%%%%%%%%%%%%%%%%
%%%%%%%%%%%%%%%%%%%%%
Before performing numerical integral, we choose the parameters 
in the FLRW universe model consistent with Planck results\cite{Ade:2015xua}, $\hat{H}_0=67.74   ~{\rm km/s/Mpc}$, $8\pi \hat{\rho}_{\rm b0}/3\hat{H}_0^2=0.04860$, 
$8\pi \hat{\rho}_{\rm d0}/3\hat{H}_0^2=0.6911$, $\hat{A}_0=2.142\times 10^{-9}$, $\hat{n}_s=0.9667$. We give the parameters $w_0$ and $w_1$ in the domain $-1.05<w_{0}<-0.95$ and $-0.1<w_{1}<0.1$.
Then, we numerically integrate Eqs.~(\ref{eq-1})--(\ref{eq-3}) under the boundary conditions 
(\ref{b-condi}), (\ref{bound_v}) and (\ref{bound_h}) and check that the obtained inhomogeneous isotropic universe model can satisfy the conditions (\ref{cl-con}) and (\ref{dz-con}) as follows.

To integrate Eqs.~(\ref{eq-1})--(\ref{eq-3}), we need to give the parameters, 
$\Omega _{\rm m}$ and $H_{0}$, of the inhomogeneous isotropic universe model.
We fix these parameters as follows.
Equation~(\ref{eq-2}) has a regular singular point at 
$z=z_{\rm cr}$ which is a root of ${\cal D}(z)=0$.
Since ${\cal D}(z)$ is a background quantity,
we can obtain $z_{\rm cr}$ without the knowledge about inhomogeneities, 
and have found  that $z_{\rm cr}$ is larger than unity for the cases of our interest.
The function ${\cal N}(z):={\cal N}(\delta t(z), v(z),z)$ should satisfy
\begin{equation}
{\cal N}(z_{\rm cr})=0 \label{regularity},
\end{equation}
so that the solutions of Eqs.~(\ref{eq-1})--(\ref{eq-3}) and their derivatives with respect to $z$ are continuous at $z=z_{\rm cr}$. 
The condition (\ref{regularity}) leads to a relation between $v|_{z=z_{\rm cr}}$, $\delta t|_{z=z_{\rm cr}}$, $\Omega_{\rm m}$ and $\bH_0$.
We assume $\Omega_{\rm m}$ and then Eq.~(\ref{regularity}) gives a relation between the remaining three quantities,  $v|_{z=z_{\rm cr}}$, $\delta t|_{z=z_{\rm cr}}$ and $H_0$.
We choose the two of three quantities and $h|_{z=z_{\rm cr}}$ so that the solutions of three differential equations (\ref{eq-1})--(\ref{eq-3}) are everywhere continuous.
We solve Eqs.~(\ref{eq-1})--(\ref{eq-3}) from $z=0$ to $z=1$ by imposing 
the boundary conditions (\ref{b-condi}), (\ref{bound_v}) and (\ref{bound_h}) and, 
at the same time, from $z=z_{\rm cr}$ to $z=1$, by making a guess at    
$\bH_0$, $h|_{z=z_{\rm cr}}$ and $v|_{z=z_{\rm cr}}$ and then fixing   
$\delta t|_{z=z_{\rm cr}}$ so that Eq.~(\ref{regularity}) is satisfied; 
If we fail to get continuous solutions of Eqs.~(\ref{eq-1})--(\ref{eq-3}), we select different 
values of $\bH_0$, $h|_{z=z_{\rm cr}}$ and $v|_{z=z_{\rm cr}}$ in accordance 
with the Newton method in the three-dimensional parameter space 
and then again integrate Eqs.~(\ref{eq-1})--(\ref{eq-3}) from 
$z=0$ to $z=1$ and, at the same time, from  $z=z_{\rm cr}$ to $z=1$; 
We iterate this procedure until the discrepancies between the values at $z=1$ obtained by the integrals from $z=0$ to $z=1$ and from $z=z_{\rm cr}$ to $z=1$ are sufficiently small;
Next, we integrate Eqs.~(\ref{eq-1})--(\ref{eq-3}) outward from $z=z_{\rm cr}$ 
with the values of $\bH_0$, $h|_{z=z_{\rm cr}}$, $v|_{z=z_{\rm cr}}$ and $\delta t|_{z=z_{\rm cr}}$
which guarantee the continuity of the solutions.
Note that Eqs.~(\ref{eq-1})--(\ref{eq-3}) implies the smoothness of solutions.
As a result, if we give $\Omega _{\rm m}$, we obtain $\bH_0, h$ and $v$  which characterize the inhomogeneous isotropic universe model in the domain $0<r<r(z=2)$ and $\bar{r}, \bar{t}, \delta r$ and $\delta t$ which characterize null geodesic in the domain $0<z<2$ from Eqs.~(\ref{eq-1})--(\ref{eq-3}).

In order to fix remaining freedoms of the inhomogeneous isotropic universe model, we use the condition (\ref{cl-con}).
If $d_{\rm A}(z_{\rm LSS})$, $\bar{\rho}_{\rm m}^{\rm LSS}$ and $\bar{\rho}_{\rm b}^{\rm LSS}$ are the same as $\hat{d}_{\rm A}(z_{\rm LSS})$, the energy density of non-relativistic matter and baryonic matter at the LSS in the FLRW universe model, respectively, the condition (\ref{cl-con}) is satisfied up to the overall factor.
We fix $A_0$ to fit the height of the first peak of the CMBR angular power spectrum 
with that of the FLRW universe model.
As a result,  $A_0$, $d_{\rm A}(z_{\rm LSS})$, $\bar{\rho}_{\rm m}^{\rm LSS}$ and $\bar{\rho}_{\rm b}^{\rm LSS}$ are uniquely determined.

\begin{figure}[htbp]
 \begin{center}
\includegraphics[width=100mm]{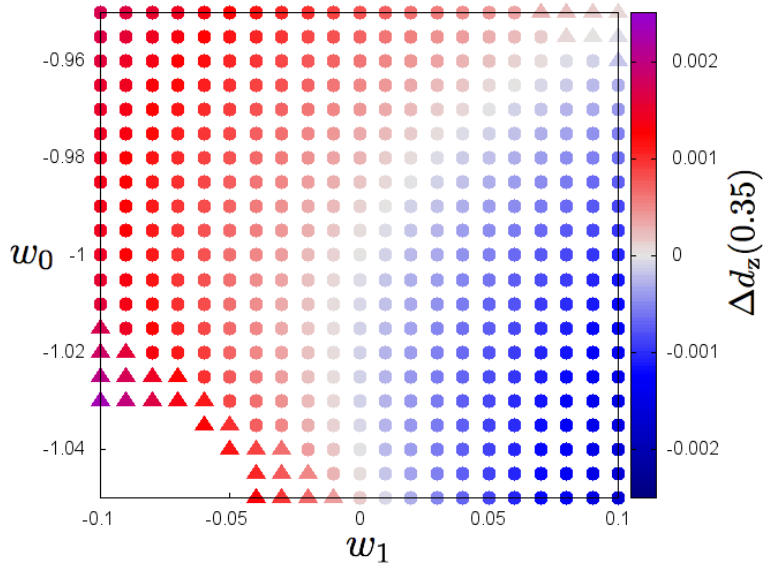}
 \end{center}
 \caption{We depict $\Delta d_{\rm z}(0.35)$ as a heat map on the $w_1$-$w_0$ plane.
The difference in the form of a point represent the difference of the density contrast $\Delta$ at the observer .
Circle means a point that absolute value of the density contrast $|\Delta|$ at the observer is less than 0.1,
whereas triangles means a point that $|\Delta|$ at the observer is more than 0.1.
The lower left corner of the blank is the domain in which the liner perturbation approximation is not vaild.}
 \label{fig:1}
\end{figure}

\begin{figure}[htbp]
 \begin{center}
\includegraphics[width=100mm]{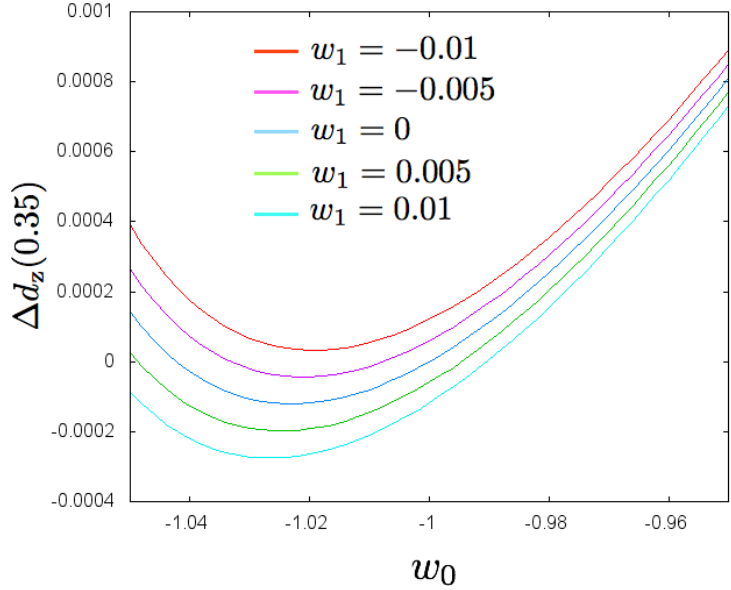}
 \end{center}
 \caption{We depict $\Delta d_{\rm z}(0.35)$ as a function of the $w_0$ with various $w_1$.}
 \label{fig:2}
\end{figure}

\begin{figure}[htbp]
 \begin{center}
\includegraphics[width=100mm]{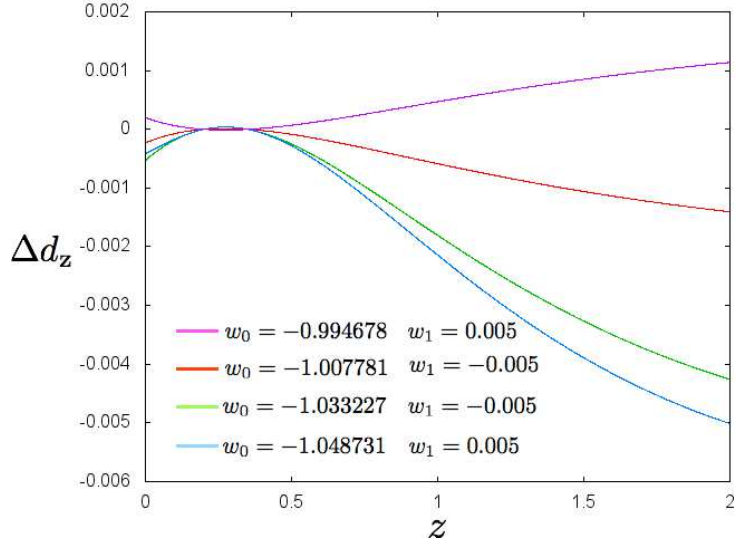}
 \end{center}
 \caption{We depict $\Delta d_{\rm z}$ as a function of $z$ in the domain $0<z<2$, for several paris of $w_0$ and $w_1$ on the curve in $w_1$-$w_0$ plane along which $\Delta d_{\rm z}(0.35)$ vanishes.}
 \label{fig:3}
\end{figure}

We check that the inhomogeneous isotropic universe model can satisfy the condition (\ref{dz-con}).
Equation (\ref{dz-inhomo}) determines $d_{\rm z}$ for arbitrary $z$.
The R.H.S of Eq.~(\ref{dz-inhomo}) is composed of $L_{\rm BAO}^{\rm (dec)}$, $\delta d_{\rm A}$, 
$\delta t$, $h$, $v$, $D_+$ and the background quantities $\bH_0$, 
$a(t_{\rm dec})$, $\bH$, $\bar{r}$ and $\bar{d}_{\rm A}$.
The background Hubble constant $\bH_0$ is determined through 
Eqs.~(\ref{eq-1})--(\ref{eq-3}) by fixing $\Omega _{\rm m}$.
The growing factor $D_+$ and the background quantities 
$\bH$, $\bar{r}$ and $\bar{d}_{\rm A}$ are completely determined by fixing $\Omega_{\rm m}$.
Equation (\ref{rho_dec}) leads to 
\begin{eqnarray}
a(t_{\rm dec})=\left( \frac{3\bH_0^2\Omega _{\rm m}}{8\pi \brho _{\rm m}(t_{\rm dec})}\right) ^{\frac{1}{3}}.
\label{adec}
\end{eqnarray} 
Since $\brho _{\rm m}(t_{\rm dec})$ is determined by the condition (\ref{cl-con}), Eq.~(\ref{adec}) 
implies that $a(t_{\rm dec})$ is also determined by  fixing $\Omega_{\rm m}$.
The perturbed angular diameter distance $\delta d_{\rm A}$ is determined by the condition (\ref{da-con}).
$L_{\rm BAO}^{\rm (dec)}$ is equal to the BAO scale at LSS , since we assume that $L_{\rm BAO}^{\rm (dec)}$ is everywhere constant, where the BAO scale at LSS is determined by the condition (\ref{cl-con}).
The remaining perturbative variables $\delta t$, $h$, $v$ are determined 
by Eqs.~(\ref{eq-1})--(\ref{eq-3}) 
once $\Omega _{\rm m}$ is fixed.
Here note that there is no condition to determine $\Omega _{\rm m}$: 
$\Omega _{\rm m}$ is still a free parameter. Hence we may rewrite $d_{\rm z}=d_{\rm z}(z)$ in the form 
\begin{eqnarray}
d_{\rm z}=d_{\rm z}(z; \Omega _{\rm m}). 
\end{eqnarray} 
Here, by using Newton method, we find a root of the one of two conditions in Eq.~(\ref{dz-con}), 
\begin{equation}
d_{\rm z}(0.2; \Omega _{\rm m})=\hat{d}_{\rm z}(0.2). \label{zero}
\end{equation}
Note that, at this stage, there is no free parameter in the inhomogeneous isotropic universe model. 
Then, we have checked whether the root of Eq.~(\ref{zero}) satisfies another one of two conditions in Eq.~(\ref{dz-con}), i.e., $d_{\rm z}(0.35; \Omega _{\rm m})=\hat{d}_{\rm z}(0.35)$.
In order to evaluate the difference of the BAO scale between the inhomogeneous isotropic universe model and the FLRW universe model, we use $\Delta d_{\rm z} (z)$ defined as
\begin{eqnarray}
\Delta d_{\rm z} (z):=\frac{d_{\rm z}(z; \Omega _{\rm m})-\hat{d}_{\rm z}(z)}{\hat{d}_{\rm z}(z)}.
\label{deltadz} 
\end{eqnarray}
In Fig.~\ref{fig:1}, we depict $\Delta d_{\rm z}(0.35)$ as a heat map on the $w_1$-$w_0$ plane.
In Fig.~\ref{fig:2}, we depict $\Delta d_{\rm z}(0.35)$ as a function of $w_0$ with various $w_1$ to see more details about $w_0$-dependence of $\Delta d_{\rm z}(0.35)$.
It can be seen from these figures that 
$\Delta d_{\rm z}(0.35)$ vanishes along a curve on $w_1$-$w_0$ plane. We denote this curve by ${\cal C}$. 
In Fig.~\ref{fig:3}, we depict $\Delta d_{\rm z}$ as a function of $z$ in the domain $0<z<2$, for several paris of $w_0$ and $w_1$ on the curve ${\cal C}$.
It can be seen from Fig.~\ref{fig:3} that the equation $d_{\rm z}(z; \Omega _{\rm m})=\hat{d}_{\rm z}(z)$ is satisfied only if $z=0.2$ or $0.35$.
Thus, if we impose the conditions (\ref{da-con})--(\ref{dz-con}), the inhomogeneous isotropic universe model can not satisfy $d_{\rm z}(z; \Omega _{\rm m})=\hat{d}_{\rm z}(z)$ at neither $z=0.2$ nor $0.35$.
This fact implies that we can, in principle, distinguish the inhomogeneous isotropic universe model 
from the FLRW universe model.
Accordingly, we can remove the systematic error, if we use distance redshift relation, the CMBR angular power spectrum and the BAO scale at three distinct redshifts as observables.

We compare the difference of the BAO scale between the inhomogeneous isotropic universe model and the FLRW universe model with the error in the observational data. 
The WiggleZ Dark Energy Survey has revealed $d_{\rm z}(0.35)=0.1097 \pm 0.0036$\cite{Blake et al.(2011)}, and the ratio of the median and the error is 0.0328.
Hence we find from Figs.~\ref{fig:1} and \ref{fig:2} that 
it is impossible to distinguish the inhomogeneous isotropic universe model 
from the FLRW universe model by using the present observational data of the BAO scale, 
as long as $-1.05<w_{0}<-0.95$ and $-0.1<w_{1}<0.1$, since the maximum value of $|\Delta d_{\rm z}(0.35)|$ is less than $2.5 \times 10^{-3}.$ 
We again consider the case that $w_0$ and $w_1$ on the curve ${\cal C}$. 
In this case, we need to compare the difference of the BAO scale between the inhomogeneous isotropic universe model and the FLRW universe model with the error in the observational data at other than $z=0.35$.
For example, we compare them at $z=0.6$.
The WiggleZ Dark Energy Survey has revealed $d_{\rm z}(0.6)=0.0726 \pm 0.0034$\cite{Blake et al.(2011)}, and the ratio of the median and the error is 0.0468.
It is larger than $|\Delta d_{\rm z}(0.6)|$ in Fig.~\ref{fig:3}, so that it is impossible to distinguish two universe models by using the present observational data.

\begin{figure}[htbp]
 \begin{center}
\includegraphics[width=100mm]{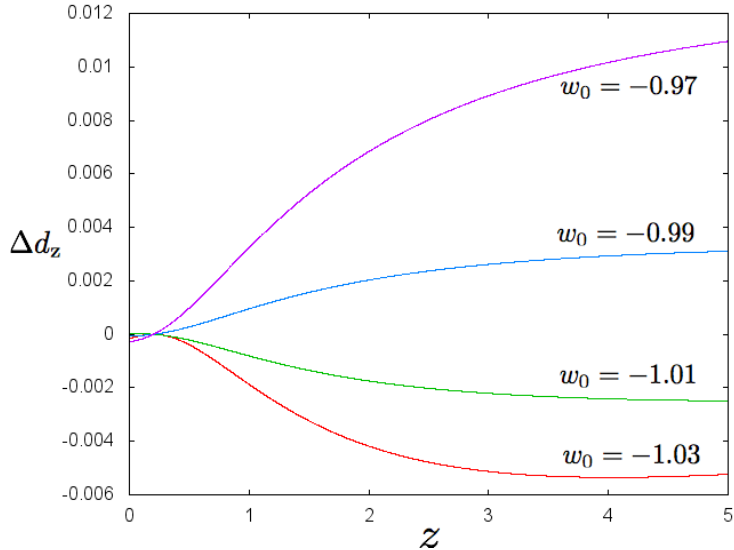}
 \end{center}
 \caption{We depict $\Delta d_{\rm z}$ as a function of the redshift z. The FLRW universe model is filled with the non-relativistic matter and the dark energy of various $w_0$ and $w_1=0$.}
 \label{fig:4}
\end{figure}

In Fig.~\ref{fig:4}, we depict $\Delta d_{\rm z}$ as a function of the redshift $z$, if the inhomogeneous isotropic universe model satisfies the conditions $d_{A}(z)=\hat{d}_{A}(z)$ in the domain $0<z<5$, Eq.~(\ref{cl-con}) and Eq.~(\ref{dz-con}) only at $z=0.2$ of the FLRW universe model with various $w_0$ of the dark energy with $w_1=0$.
It can be seen from this figure that the larger $z$ is, the larger $|\Delta d_{\rm z}|$ becomes, if $z$ is larger that 0.2.
Thus, it is important to get observational data of the distance-redshift relation and the BAO scale at $z$ larger than unity.

%%%%%%%%%%%%%%%%%%%%%
%%%%%%%%%%%%%%%%%%%%%
\section{Summary and discussion}
%%%%%%%%%%%%%%%%%%%%%
%%%%%%%%%%%%%%%%%%%%%
We studied whether we can distinguish the inhomogeneous isotropic universe model with the cosmological constant from the FLRW universe model with dark energy other than the cosmological constant and remove the systematic error due to isotropic inhomogeneities, by considering multiple observables: the distance-redshift relation, the fluctuation spectrum of the CMBR and the BAO scale. We found that we can do so; There 
is no inhomogeneous isotropic universe model whose distance-redshift relation, 
fluctuation spectrum of the CMBR and the BAO scale are identical with those of the FLRW universe model, 
as long as the density perturbations are adiabatic.
It is nontrivial that we can distinguish the inhomogeneous isotropic universe model from the FLRW universe model by using a finite number of observables, since the inhomogeneous isotropic universe model has a functional degree of freedom.

Here it should be noted that we have used not only the information about 
the universe on the past light cone but also that 
inside the past light cone, since we assumed that the BAO scale at the decoupling time is everywhere constant.
However, there is a possibility that the BAO scale at the decoupling time is inhomogeneous, if the ratio between the energy densities of baryonic matter and CDM has been inhomogeneous at the decoupling time.
If the ratio between the energy densities of baryonic matter and CDM is inhomogeneous and isotropic, we can not distinguish the inhomogeneous isotropic universe model from the FLRW universe model by only the distance-redshift relation, the fluctuation spectrum of the CMBR and the BAO scale. In the case of  
$w_{0}=-1.01$ and $w_{1}=0$, the inhomogeneous isotropic universe model can also  
explain observations, if the fluctuation of the ratio between the energy densities of baryonic matter 
and CDM at $t=t_0$ is $-0.037408$ (see Appendix A).
It is very important to observe the ratio between the energy densities of baryonic matter 
and CDM in the domain $0<z<2$.

\section*{Acknowledgments}
We are grateful to Chul-Moon Yoo, Ryusuke Nishikawa, Hideki Ishihara and colleagues in the group of elementary particle physics and gravity at Osaka City University for useful discussions and helpful comments.
K. N. was supported in part by JSPS KAKENHI Grant No. 25400265.

%Appendix
\appendix
\section{The ratio of baryonic matter and CDM}
Here, we discuss whether the inhomogeneous isotropic universe model can satisfy the all of the conditions (\ref{da-con})--(\ref{dz-con}), if the ratio between the energy densities of baryonic matter and CDM depends on $r$.
We assume that the comoving length scale of fluctuation of ratio between the energy densities of baryonic matter and CDM is the order of that of the present horizon and the energy density of radiation $\rho _{\rm \gamma} (t, r)$ does not depend on $r$.

The energy densities of non-relativistic matter $\rho _{\rm m}(t, r)$ and baryonic matter $\rho _{\rm b}(t, r)$ are write in the form,
\begin{eqnarray}
\rho _{\rm m}(t, r)=\bar{\rho} _{\rm m}(1+ \Delta (t, r)),
\label{baryon} 
\end{eqnarray} 
\begin{eqnarray}
\rho _{\rm b} (t, r) =\bar{\rho} _{\rm b}(1+ \Delta _{\rm b}(t, r)),
\label{baryon} 
\end{eqnarray} 
where $\Delta _{\rm b}(t, r)$ is the density contrast of baryonic matter. 
For simplicity, we assume that 
\begin{eqnarray}
\Delta _{\rm b}(t_0, r)|_{r=r(z=0.2)} = \Delta _{\rm b}(t_0, r) |_{r=r(z=0.35)}=\delta _{\rm b}.
\label{delta-dec}
\end{eqnarray} 
where $\delta _{\rm b}$ is arbitrary constant.

The condition (\ref{da-con}) does not impose any constraints on the ratio between the energy densities of baryonic matter and CDM.
By the assumption, the ratio between the energy densities of baryonic matter and CDM is almost spatially constant in the vicinity of the LSS, so that the condition (\ref{cl-con}) fixes 
$A_0, d_{\rm A}(z_{\rm LSS}), \bar{\rho}_{\rm m}^{\rm LSS}$ and $\bar{\rho}_{\rm b}^{\rm LSS}$ in the same way as in the case that the ratio between the energy densities of baryonic matter and CDM does not fluctuate.
By contrast, the fluctuation of the ratio between the energy densities of baryonic matter and CDM affects the sound velocity of baryonic matter, hence $L_{\rm BAO}^{\rm (dec)}$ may depend on $r$.
Since we consider the inhomogeneous isotropic universe model whose comoving length scale of inhomogeneities is comparable to that of the present horizon, 
$L_{\rm BAO}^{\rm (dec)}(r)$ can be obtained by regarding our inhomogeneous isotropic universe model as a homogenous isotropic universe model in each domain of BAO scale at decoupling time.
Hence $L_{\rm BAO}^{\rm (dec)}$ at each $r$ is approximately determined by the fitting formulae given in Ref.~\cite{Eisenstein:1997ik}, since we have assumed that $\rho _{\gamma}$ at the decoupling time is equal to the value at the LSS.

Here note that $\delta _{\rm b}$ is a free parameter.
We may choose $\delta _{\rm b}$ so that the inhomogeneous isotropic universe model satisfies the conditions (\ref{da-con})--(\ref{dz-con}).
For example, in the case of the FLRW universe model with $w_{0}=-1.01$ and $w_1=0$,  
the inhomogeneous isotropic universe model satisfies the conditions (\ref{da-con})--(\ref{dz-con}), 
if we assume $\delta _{\rm b}=-0.017950$.
The fluctuation of the ratio between the energy densities of baryonic matter and CDM
\begin{eqnarray}
\frac{\displaystyle{\frac{\rho _{\rm b}(t_0, r|_{z=0.35})}{\rho _{\rm CDM}(t_0, r|_{z=0.35})}}-\frac{\bar{\rho} _{\rm b}(t_0)}{\bar{\rho} _{\rm CDM}(t_0)}
}{\displaystyle{\frac{\bar{\rho} _{\rm b}(t_0)}{\bar{\rho} _{\rm CDM}(t_0)}}}=-0.037408,
\label{b_m}
\end{eqnarray}

%references

\end{document}